\documentclass[preprint]{elsarticle}

\usepackage{amsmath,amsfonts,amssymb}
\usepackage{latexsym}
\usepackage{dcolumn}
\usepackage{graphicx,epsfig}
\usepackage{amsthm}
\usepackage{hyperref}

\begin{document}
\title{\bf {Black Hole Entropy and the Modified Uncertainty Principle: A heuristic analysis}}
\author{Barun Majumder}
\ead{barunbasanta@iiserkol.ac.in}
\address{Indian Institute of Science Education and Research (Kolkata) \\ Mohanpur, Nadia, West Bengal, Pin 741252\\ India}
\begin{frontmatter}
\begin{abstract} 
Recently Ali \textit{et.al.} (2009) proposed a Generalized Uncertainty Principle (or GUP) with a linear term in momentum (accompanied by Plank length). Inspired
by this idea here we calculate the quantum corrected value of a Schwarzschild black hole entropy and a Reissner-Nordstr$\ddot{o}$m black hole with double
horizon by utilizing the proposed generalized uncertainty principle. We find that the leading order correction goes with the square root of the horizon area
contributing positively. We also find that the prefactor of the logarithmic contribution is negative and the value exactly matches with some earlier existing calculations. With the Reissner-Nordstr$\ddot{o}$m black hole we see that this model independent procedure is not only valid for single horizon spacetime
but also valid for spacetimes with inner and outer horizons.
\end{abstract}
\end{frontmatter}


The realization that black holes are thermodynamic objects with well defined entropy and temperature is one of the landmark achievement in theoretical
physics \cite{w1,w2,w3}. Hawking \cite{w3} has shown that a Schwarzschild black hole has a thermal radiation with a temperature $T_H=\frac{1}{8\pi M}$,
where $M$ is the mass of the black hole. Also the entropy associated with a Schwarzschild black hole is given by the Bekenstein-Hawking entropy-area
relation
\begin{equation}
S_{BH} = \frac{A}{4l_p^2}~~.
\end{equation} 
Here $A$ is the cross sectional area of the black hole horizon. Recently there has been much attention devoted to resolving the quantum corrections to
the black hole entropy. As entropy has a definite statistical meaning in the thermodynamic system, it accounts for the number of microstates of the system. A
thermodynamic system is composed of atoms and molecules but nothing in particular can be said about the black hole except the presence of strong gravity. It
is now common in literature that black hole entropy can be attributed a definite statistical meaning (though this belief warrants a certain degree of caution \cite{w4}). The main problem in the study of black hole entropy is to identify the microstates and count them. Two leading candidate theory of quantum gravity
(aimed for a successful quantum theory of gravity) namely, string theory and loop quantum gravity, both achieved an enormous amount of success in statistical
explanation of the entropy-area law (we can see \cite{w5,w6} for a brief overview). In this discussion we will mainly focus on the quantum-corrected entropy.
Various theories of quantum gravity (e.g.,\cite{w7,w8,w9,x1,w10}) have predicted the following expansive form:
\begin{equation}
\label{e2}
S= \frac{A}{4l_p^2} ~+~ c_0~\ln \bigg(\frac{A}{4l_p^2}\bigg) ~+~ \sum_{n=1}^\infty ~c_n \bigg(\frac{A}{4l_p^2}\bigg)^{-n} ~+~ \mathit{const.}~~,
\end{equation}
where the coefficients $c_n$ can be regarded as model dependent parameters. Many researchers have expressed a vested interest in fixing $c_0$ (the
coefficient of the subleading logarithmic term) \cite{w7}. Recent rigorous calculations of loop quantum gravity predicts the value of $c_0$ to be $-1/2$ \cite{w10}. 
\par
For the study of black hole entropy we can also use a model independent concept namely the Generalized Uncertainty Principle or GUP. The idea that the uncertainty
principle could be affected by gravity was given by Mead \cite{w11}. In the regime when the gravity is strong enough, conventional Heisenberg uncertainty
relation is no longer satisfactory (though approximately but perfectly valid in low gravity regimes). Later modified commutation relations between position and momenta commonly known as Generalized Uncertainty Principle were given by candidate theories of quantum gravity (String Theory, Doubly Special
Relativity (or DSR) Theory and Black Hole Physics) with the prediction of a minimum measurable length \cite{b1,b2,b3}. Similar kind of
commutation relation can also be found in the context of Polymer Quantization in terms of polymer mass scale \cite{b4}. Importance of the GUP can also be realized
on the basis of simple \textit{gedanken} experiments without any reference of a particular fundamental theory \cite{b2}. So we can think the GUP as a model
independent concept, ideally perfect for the study of black hole entropy. Many authors have applied the GUP for a heuristic analysis of the black hole
entropy (we can see \cite{w8,x2,w12,w13} for a brief idea). 
\par
The authors in \cite{b5} proposed a GUP which is consistent with DSR theory, string theory and black hole physics and which says
\begin{equation}
\left[x_i,x_j\right] = \left[p_i,p_j\right] = 0 ~~,
\end{equation}
\begin{equation}
\label{e4}
[x_i, p_j] = i \hbar \left[  \delta_{ij} -  l  \left( p \delta_{ij} + \frac{p_i p_j}{p} \right) + l^2  \left( p^2 \delta_{ij}  + 3 p_{i} p_{j} \right) \right]~~,
\end{equation}
\begin{align}
\label{e5}
 \delta x ~\delta p &\geq \frac{\hbar}{2} \left[ 1 - 2 l \langle p \rangle + 4 l^2 \langle p^2\rangle \right]  \nonumber \\
& \geq \frac{\hbar}{2} \left[ 1  +  \left(\frac{l}{\sqrt{\langle p^2 \rangle}} + 4 l^2  \right)  (\delta p)^2  +  4 l^2 \langle p \rangle^2 -  2 l \sqrt{\langle p^2 \rangle}~ \right],
\end{align}
where $ l=\frac{l_0 l_{p}}{\hbar} $. Here $ l_{p} $ is the Plank length ($ \approx 10^{-35} m $). It is normally assumed that the dimensionless
parameter $l_0$ is of the order unity. If this is the case then the $l$ dependent terms are only important at or near the Plank
regime. But here we expect the existence of a new intermediate physical length scale of the order of $l \hbar = l_0 l_{p}$. We also note
that this unobserved length scale cannot exceed the electroweak length scale \cite{b5} which implies $l_0 \leq 10^{17}$. These equations are
approximately covariant under DSR transformations but not Lorentz covariant \cite{b3}. These equations also imply
\begin{equation}
\delta x \geq \left(\delta x \right)_{min} \approx l_0\,l_{p}
\end{equation}
and
\begin{equation}
\delta p \leq \left(\delta p \right)_{max} \approx \frac{M_{p}c}{l_0}
\end{equation}
where $ M_{p} $ is the Plank mass and $c$ is the velocity of light in vacuum. With a lower bound for position fluctuations we can claim that there is a minimum measurable distance and from an upper bound of momentum fluctuations we claim that momentum measurements cannot be arbitrarily imprecise. The effect of this proposed GUP is well studied recently for some well known physical systems in \cite{b5,b6,w17}.
\par
In this Letter we apply this newly proposed GUP for a perturbative calculation of the quantum-corrected entropy which can be readily extended to any desired order. In the first half we consider a Schwarzschild black hole. In the next half we do the same analysis for a Reissner-Nordstr$\ddot{o}$m spacetime with double horizons.
\par
Equations (\ref{e4}) and (\ref{e5}) represents modified Heisenberg algebra. But the interesting part of these two relation is the term which is linear
in $l(=l_0l_p/\hbar)$ with $p$. Inspired by this idea, for our purpose we will consider the modified Heisenberg algebra (modified Heisenberg principle) with
a small change in notation where $x$ and $p$ obeys the relation ($\alpha >0$)
\begin{equation}
\label{e8}
\delta x ~\delta p \geq \hbar~ \bigg[~1 - \frac{\alpha l_p}{\hbar} ~\delta p + \frac{\alpha^2 l_p^2}{\hbar^2} ~(\delta p)^2~\bigg] ~~.   
\end{equation}  
In writing equation (\ref{e8}) we made an approximation that $(\delta p)\approx \sqrt{\langle p^2 \rangle}$. This means $\langle p \rangle \approx 0$. Now
this seems to be a valid approximation as we are going to study the Schwarzschild black hole which is spherically symmetric \footnote{Also in many problems of
usual quantum mechanics we find $\langle p \rangle =\langle x \rangle =0~$ (for ex. ground state of harmonic oscillator).}.  
We can see that if $\alpha =2l_0$ this is the same relation as that of (\ref{e5}). Here $\delta x$ and $\delta p$ are the position and momentum uncertainty
for a quantum particle and $\alpha$ is a dimensionless positive parameter (also known as deformation parameter in the literature of non-commutative geometry).
As $l_p=\sqrt{\frac{\hbar G}{c^3}}$, where $G$ is the Newtonian coupling constant, we can imply that the extra terms in the uncertainty relation is a consequence
of gravity. We can re express the modified Heisenberg principle (or MUP) of (\ref{e8}) in the following form
\begin{equation}
\label{e9}
\delta p \geq \frac{~\hbar~(\delta x + \alpha l_p) ~-~ \hbar \sqrt{(\delta x + \alpha l_p)^2 - 4\alpha^2 l_p^2}~}{2\alpha^2 l_p^2}~~,
\end{equation}
where a negative sign choice is made by taking the classical limit. As $l_p$ is normally viewed as an ultraviolet cut-off on spacetime geometry (e.g.,\cite{w14}),
it is quite justified that we can consider the dimensionless ratio $\frac{l_p}{\delta x}$ relatively small as compared to unity. So we can Taylor expand
equation (\ref{e9}) and rewrite the same equation after some simple manipulation as
\begin{equation}
\delta p \geq \frac{1}{\delta x} \bigg[1 - \frac{\alpha l_p}{2(\delta x)} + \frac{\alpha^2 l_p^2}{2(\delta x)^2} - \frac{\alpha^3 l_p^3}{2(\delta x)^3}
+ \frac{9}{16}\frac{\alpha^4 l_p^4}{(\delta x)^4} - \ldots \bigg] ~~,
\end{equation}
where we have considered a choice of unit with $\hbar =1$. The Heisenberg uncertainty principle ($\delta p \delta x \geq 1$) can be translated to the lower bound
$E\delta x \geq 1$ with the arguments used in \cite{w15,x1}, where $E$ is the energy of a quantum particle. The measurement process considered here uses a photon
to specify the position of the quantum particle. If we imply our MUP, we can rebuild the lower bound as
\begin{equation}
E \geq \frac{1}{\delta x} \bigg[1 - \frac{\alpha l_p}{2(\delta x)} + \frac{\alpha^2 l_p^2}{2(\delta x)^2} - \frac{\alpha^3 l_p^3}{2(\delta x)^3}
+ \frac{9}{16}\frac{\alpha^4 l_p^4}{(\delta x)^4} - \ldots \bigg] ~~.
\end{equation}
Now we will consider the picture where a quantum particle in the immediate vicinity of an event horizon is absorbed by the black hole. From the knowledge
of general relativity we know that for a black hole, absorbing a classical particle with energy $E$ and size $R$, the minimum increase in area is expressed as
\begin{equation}
\label{e12}
\Delta A_{min} \geq 8 \pi l_p^2 E R ~~.
\end{equation}
For a quantum particle $R$ can never be less than the intrinsic uncertainty in the position of the particle \cite{w2}. Hence for a quantum particle
equation (\ref{e12}) reduces to 
\begin{equation}
\label{e13}
\Delta A_{min} \geq 8 \pi l_p^2 E \delta x ~~.
\end{equation}
Considering MUP we can re express equation (\ref{e13}) as 
\begin{equation}
\label{e14}
\Delta A_{min} \simeq \epsilon ~l_p^2 ~\bigg[1 - \frac{\alpha l_p}{2(\delta x)} + \frac{\alpha^2 l_p^2}{2(\delta x)^2} - \frac{\alpha^3 l_p^3}{2(\delta x)^3}
+ \frac{9}{16}\frac{\alpha^4 l_p^4}{(\delta x)^4} - \ldots \bigg] ~~.
\end{equation}
Here $\epsilon$ is a numerical factor greater than the order of $8\pi$. 
\par
Let us now consider a Schwarzschild black hole of constant mass immersed in a bath of radiation in its own temperature. So the framework is in principle
\textit{microcanonical}. The particles considered in the last section should have a Compton wave length of the order of the inverse of the Hawking
temperature \cite{w3}. Usually the inverse of surface gravity is the best choice of length scale near horizon. Here also we will
choose (we can see \cite{w16,x2} for a brief argument)
\begin{equation}
\delta x \sim 2r_s ~~.
\end{equation}
Identifying $\delta x \sim \sqrt{\frac{A}{\pi}}$ and putting this in equation (\ref{e14}) we get 
\begin{equation}
\Delta A_{min} \simeq \epsilon ~l_p^2~\bigg[1 - \frac{\alpha ~l_p ~\pi^{1/2}}{2~A^{1/2}} + \frac{\alpha^2 ~l_p^2 ~\pi}{2~A} - 
\frac{\alpha^3 ~l_p^3 ~\pi^{3/2}}{2~A^{3/2}} + \frac{9~\alpha^4 ~l_p^4 ~\pi^{2}}{16~A^{2}} - \ldots \bigg] ~~.
\end{equation}
Bekenstein first argued \cite{w2} that the black hole entropy should depend on the horizon area. Also the minimum increase of entropy is one \textit{bit}
$b$ and generally it is considered that $b=\ln 2$. Using this we now write
\begin{equation}
\frac{d S}{d A} \simeq \frac{\Delta S_{min}}{\Delta A_{min}} \simeq \frac{b}{\epsilon ~l_p^2~\bigg[1 - \frac{\alpha ~l_p ~\pi^{1/2}}{2~A^{1/2}} + \frac{\alpha^2 ~l_p^2 ~\pi}{2~A} - \frac{\alpha^3 ~l_p^3 ~\pi^{3/2}}{2~A^{3/2}} + \frac{9~\alpha^4 ~l_p^4 ~\pi^{2}}{16~A^{2}} - \ldots \bigg]} ~~.
\end{equation}
Following the same procedure as before (for performing Taylor expansion) we write the same equation as 
\begin{equation}
\frac{d S}{d A} \simeq \frac{b}{\epsilon ~l_p^2} ~\bigg[1 + \frac{\alpha ~l_p ~\pi^{1/2}}{2~A^{1/2}} - \frac{\alpha^2 ~l_p^2 ~\pi}{4~A} + \frac{\alpha^3 ~l_p^3 ~\pi^{3/2}}{8~A^{3/2}} - \frac{\alpha^4 ~l_p^4 ~\pi^{2}}{8~A^{2}} + \ldots \bigg] ~~.
\end{equation}
Integrating we get (up to an additive constant factor of integration)
\begin{equation}
\label{e19}
S \simeq \frac{A}{4~l_p^2} + \frac{\pi^{1/2} ~\alpha}{2}~\sqrt{\frac{A}{4~l_p^2}} - \frac{\pi ~\alpha^2}{16} \ln \frac{A}{4~l_p^2} - 
\frac{\pi^{3/2} ~\alpha^3}{32} \bigg(\frac{A}{4~l_p^2}\bigg)^{-1/2} + \frac{\pi^2 ~\alpha^4}{128} \bigg(\frac{A}{4~l_p^2}\bigg)^{-1} - 
\ldots + \mathit{const.} ~~.
\end{equation}
Here we have compared the first term with Bekenstein-Hawking entropy-area relation which says $b/\epsilon$ should be $1/4$. Equation (\ref{e19}) can
be written in the form of an expansion
\begin{equation}
S \simeq \frac{A}{4~l_p^2} + \frac{\pi^{1/2} ~\alpha}{2}~\sqrt{\frac{A}{4~l_p^2}} - \frac{\pi ~\alpha^2}{16} \ln \frac{A}{4~l_p^2} - 
\sum_{m=\frac{1}{2},\frac{3}{2},~\ldots}^\infty d_m ~ \bigg(\frac{A}{4~l_p^2}\bigg)^{-m} + 
\sum_{n=1,2,~\ldots}^\infty c_n ~ \bigg(\frac{A}{4~l_p^2}\bigg)^{-n} + \mathit{const.} ~~.
\end{equation}
Here $m$ denotes positive half-integers and $n$ positive integers. If we compare this equation with (\ref{e2}) we can see that there are extra terms in this equation. One of the leading contribution to the entropy is from the new second term $\sim \sqrt{\mathit{Area}}$. In the context of
equation (\ref{e4}) and (\ref{e5}) this was first pointed out in \cite{w17}.
Also we have terms proportional to $(\mathit{Area})^{-m}$. This is a consequence of the form of the modified uncertainty relation which we have
used\footnote{The hypothesis of modified energy-momentum dispersion relation (commonly known as MDR) is popular among those adopting a \textit{spacetime foam} intuition in the study of the quantum gravity problem. In most cases one is led to consider a dispersion relation of the type
\begin{equation}
p^2 \simeq E^2 - \mu^2 + \alpha_1 l_p E^3 + \alpha_2 l_p^2 E^4 + \dots ~~, \nonumber
\end{equation}
where $\mu$ is termed as mass parameter and it is directly related to the rest energy of the particle. This type of modified dispersion relations are used to evaluate black hole entropy. If the cubic term $\alpha_1l_pE^3$ is present in the energy-momentum dispersion relation then the leading correction
goes like $\sqrt{\mathit{Area}}$. For a brief discussion we can see \cite{w13}.}.
A linear term in Plank length accompanied by $p$ in the commutation relation of $x$ and $p$ gives this new contribution to the quantum corrected
entropy-area relation. 
\par
Let us now consider the case of a Reissner-Nordstr$\ddot{o}$m black hole with double horizon. The line element of this spacetime is given by
\begin{equation}
d s^2 = -\bigg(1 - \frac{2M}{r} + \frac{Q^2}{r^2}\bigg)~dt^2 + \bigg(1 - \frac{2M}{r} + \frac{Q^2}{r^2}\bigg)^{-1} dr^2 + r^2 ~d \Omega_2^2 ~~,
\end{equation}
where $r_\pm = M \pm \sqrt{M^2 - Q^2}$ are the locations of outer and inner horizons. $Q$ is the electric charge of the black hole and we will consider it as constant. For simplicity we have considered the choice of unit where $G=c=1$. Using similar arguments as in the case of a Schwarzschild black hole we will
consider 
\begin{equation}
\delta x \sim 2(r_+ - r_-) ~~.
\end{equation}
With a simple manipulation we can write this as 
\begin{equation}
\delta x \sim \sqrt{\frac{A}{\pi}}~\bigg(1 - \frac{4~\pi ~Q^2}{A}\bigg) ~~.
\end{equation}
Here the minimum increase in the horizon area can be expressed as
\begin{align}
\Delta A_{min} \simeq &\epsilon ~l_p^2~\bigg[1 - \frac{\alpha ~l_p ~\pi^{1/2}}{2~A^{1/2}}~ \bigg(1 - \frac{4~\pi~Q^2}{A}\bigg)^{-1} + 
\frac{\alpha^2 ~l_p^2 ~\pi}{2~A}~ \bigg(1 - \frac{4~\pi~Q^2}{A}\bigg)^{-2} \nonumber \\
& - \frac{\alpha^3 ~l_p^3 ~\pi^{3/2}}{2~A^{3/2}}~ \bigg(1 - \frac{4~\pi~Q^2}{A}\bigg)^{-3} + 
\frac{9~\alpha^4 ~l_p^4 ~\pi^{2}}{16~A^{2}}~ \bigg(1 - \frac{4~\pi~Q^2}{A}\bigg)^{-4} - \ldots \bigg] ~~.
\end{align}
We now write the differential entropy-area relation for this black hole as
\begin{align}
\frac{d S}{d A} \simeq & \frac{b}{\epsilon ~l_p^2} ~\bigg[1 + \frac{\alpha ~l_p ~ \pi^{1/2}}{2~A^{1/2}}~\bigg(1 + \frac{4~\pi ~Q^2}{A} + 
\frac{16~\pi^2 ~Q^4}{A^2}~\bigg) - \frac{\alpha^2 ~l_p^2 ~ \pi}{4~A}~\bigg(1 + \frac{8~\pi ~Q^2}{A} + \frac{48~\pi^2 ~Q^4}{A^2}~\bigg) \nonumber \\
& + \frac{\alpha^3 ~l_p^3 ~ \pi^{3/2}}{8~A^{3/2}}~\bigg(1 + \frac{12~\pi ~Q^2}{A} + \frac{96~\pi^2 ~Q^4}{A^2}~\bigg) - \frac{\alpha^4 ~l_p^4 ~ \pi^{2}}{8~A^{2}}~\bigg(1 + \frac{16~\pi ~Q^2}{A} + \frac{160~\pi^2 ~Q^4}{A^2}~\bigg) + \ldots \bigg] ~~.
\end{align}
Here we have considered terms up to the ${\cal O}(\frac{Q^4}{A^2})$ while performing the Taylor expansion. Higher order terms are neglected with the assumption
$A>>Q$. With the calibrated value of $b/\epsilon = 1/4$, the final expression of the quantum corrected entropy-area relation (up to an additive constant factor
of integration) for the Reissner-Nordstr$\ddot{o}$m black hole with double horizon is written as
\begin{align}
S \simeq & \frac{A}{4~l_p^2} + \frac{\alpha ~\pi^{1/2}}{4~l_p}~\bigg(A^{1/2} - \frac{4~\pi ~Q^2}{A^{1/2}} - \frac{16~\pi^2 ~Q^4}{3~A^{3/2}}\bigg) - 
\frac{\alpha^2 ~\pi}{16}~\bigg(\ln A - \frac{8~\pi ~Q^2}{A} - \frac{24~\pi^2 ~Q^4}{A^{2}}\bigg) \nonumber \\
& - \frac{\alpha^3 ~l_p^3~\pi^{3/2}}{16}~\bigg(\frac{1}{A^{1/2}} + \frac{4~\pi ~Q^2}{A^{3/2}} + \frac{96~\pi^2 ~Q^4}{5~A^{5/2}}\bigg) +
\frac{\alpha^4 ~l_p^2~\pi^{2}}{32}~\bigg(\frac{1}{A} + \frac{8~\pi ~Q^2}{A^{2}} + \frac{160~\pi^2 ~Q^4}{3~A^{3}}\bigg) - \ldots + \mathit{const.} ~~. 
\end{align}
Clearly we can see the if $Q=0$, we get back equation (\ref{e19}).
\par
So in this Letter, we have exploited the generalized uncertainty principle as proposed by Ali \textit{et. al.} \cite{b5} to evaluate the quantum
corrected black hole entropy for a Schwarzschild black hole and a Reissner-Nordstr$\ddot{o}$m black hole with double horizon. We found that the leading order correction to the Bekenstein-Hawking entropy-area relation goes as $\mathit{\sqrt{Area}}$ contributing positively. This term can also be obtained if we use the modified energy-momentum dispersion relation containing a term proportional to $l_p\times \mathit{(Energy)^3}$ for the calculation of black hole entropy.
Some models of quantum gravity disregard this term. The next leading order contribution to the entropy goes as the logarithm of the area but it contributes negatively. Apart from these two corrections we also found two series expansion which goes with the negative power of \textit{Area}. One series is consistent
with the calculation performed with the earlier version of GUP (e.g.,\cite{w8}). The other one goes with negative half-integer powers
of \textit{Area} contributing negatively to the entropy. Here we have found that the logarithmic prefactor takes on the value $-\frac{\pi \alpha^2}{16}$.
If we look back at equation (\ref{e5}) this value is $-\frac{\pi l_0^2}{4}$ (according to the newly proposed GUP \cite{b5}). Though we are unable to
make a precise statement about $l_0$ but still this is exactly the same value as deduced by the authors in \cite{w19,w8}. We have considered that the
black hole is immersed in a bath of radiation at its own temperature, hence we have computed the \textit{microcanonical} entropy. Later we have
utilized the same procedure for a Reissner-Nordstr$\ddot{o}$m black hole with double horizon. We also found that this procedure as
mentioned in \cite{w18} is not only valid for single horizon spacetime but also valid for spacetimes with outer and inner horizons.


\section*{Acknowledgements}
The author is very much thankful to Prof. Narayan Banerjee for helpful discussions and guidance. The author would also like to
thank an anonymous referee for helpful comments and enlightening suggestions.

\end{document}